\documentclass[%
 reprint,
superscriptaddress,
nofootinbib,
 amsmath,amssymb,
 aps,
prl,
]{revtex4-2}

\usepackage{graphicx}
\usepackage{dcolumn}
\usepackage{bm}
\usepackage{xcolor}
\usepackage{float}	
\usepackage{soul}
\usepackage{url}
\usepackage{siunitx}

\begin{document}

\newcommand{\bra}[1]{\left\langle\,#1\,\right|}
\newcommand{\ket}[1]{\left|\,#1\,\right\rangle}

\newcommand{\pb}{PtBi$_2$}

\newcommand{\jf}[1]{\textcolor{blue}{#1}}
\newcommand{\rjf}[2]{\textcolor{red}{\st{#1}}\textcolor{blue}{#2}}
\newcommand{\sjf}[1]{\textcolor{red}{\st{#1}}}

\preprint{APS/123-QED}

\title{Disentangling bulk and surface states in the electronic structure of PtBi$_2$(0001)}

\author {Stefanie~Suzanne~Brinkman}
\email{stefanie.brinkman@ntnu.no}
\affiliation {Center for Quantum Spintronics, Department of Physics, Norwegian University of Science and Technology, 7491 Trondheim, Norway}

\author {Xin~Liang~Tan}
\affiliation {Center for Quantum Spintronics, Department of Physics, Norwegian University of Science and Technology, 7491 Trondheim, Norway}

\author {Anders~Christian~Mathisen}
\affiliation {Center for Quantum Spintronics, Department of Physics, Norwegian University of Science and Technology, 7491 Trondheim, Norway}

\author {Fabian G\"ohler}
\affiliation {Center for Quantum Spintronics, Department of Physics, Norwegian University of Science and Technology, 7491 Trondheim, Norway}

\author {\O yvind~Finnseth}
\affiliation {Department of Materials Science and Engineering, Norwegian University of Science and Technology, 7491 Trondheim, Norway}

\author {Chul-Hee Min}
\affiliation {Center for Quantum Spintronics, Department of Physics, Norwegian University of Science and Technology, 7491 Trondheim, Norway}

\author{Grigory Shipunov}
\affiliation{Van der Waals-Zeeman Institute, Department of Physics and Astronomy,
University of Amsterdam, Science Park 904, 1098 XH Amsterdam, The Netherlands}

\author{Falk Pabst}
\affiliation{Van der Waals-Zeeman Institute, Department of Physics and Astronomy,
University of Amsterdam, Science Park 904, 1098 XH Amsterdam, The Netherlands}

\author{Manuel Alonso Lemos}
\affiliation{Instituto Balseiro, Univ. Nacional de Cuyo, Av. Bustillo 9500, Argentina}

\author {Balasubramanian Thiagarajan}
\affiliation {MAX IV Laboratory, Lund University, Lund, Sweden}

\author {Craig Polley}
\affiliation {MAX IV Laboratory, Lund University, Lund, Sweden}

\author {Masashi Arita}
\affiliation {Research Institute for Synchrotron Radiation Science (HiSOR), Hiroshima University, Higashi-Hiroshima, Japan}

\author {Kenya Shimada}
\affiliation {Research Institute for Synchrotron Radiation Science (HiSOR), Hiroshima University, Higashi-Hiroshima, Japan}
\affiliation{International Institute for Sustainability with Knotted Chiral Meta Matter (WPI-SKCM$^2$), Hiroshima University, Higashi-Hiroshima, Japan}
\affiliation{Research Institute for Semiconductor Engineering (RISE), Hiroshima University, Higashi-Hiroshima, Japan}

\author{Anna Isaeva}
\affiliation{Faculty of Physics, Technical University of Dortmund, Otto-Hahn-Straße 4, D-44227 Dortmund, Germany}
\affiliation{Center Future Energy Materials and Systems (RC FEMS), Germany}\affiliation{Van der Waals-Zeeman Institute, Department of Physics and Astronomy,
University of Amsterdam, Science Park 904, 1098 XH Amsterdam, The Netherlands}

\author{Jorge I. Facio}
\affiliation{Instituto Balseiro, Univ. Nacional de Cuyo, Av. Bustillo 9500, Argentina}
\affiliation{Centro Atómico Bariloche, Instituto de Nanociencia y Nanotecnología (CNEA-CONICET), Av. Bustillo 9500, Argentina}

\author {Hendrik~Bentmann}
\affiliation {Center for Quantum Spintronics, Department of Physics, Norwegian University of Science and Technology, 7491 Trondheim, Norway}

\date{\today}

\begin{abstract} 
Recent reports of surface-localized topological superconductivity in trigonal PtBi$_2$ highlight the importance of understanding its surface electronic structure. We investigate the bulk and surface band structure of PtBi$_2$ using angle-resolved photoemission spectroscopy (ARPES) and first-principles calculations. Through photon-energy- and polarization-dependent measurements, we disentangle bulk dispersions from surface states on the two distinct surface terminations of PtBi$_2$(0001). For both terminations, we assign several different surface states and find good agreement between experiment and calculations. Based on our calculations, we analyze the orbital composition in the surface and bulk bands and compare the results to polarization-dependent ARPES measurements. Together, our results provide a coherent picture of the surface electronic structure of PtBi$_2$ across both surface terminations.

\end{abstract}

\maketitle 

\section{Introduction}
\label{intro}

Topological superconductors have attracted attention because they are expected to host Majorana quasiparticles, making them a potential platform for topological quantum computation \cite{Nayak:08}. Weyl semimetals are candidates for realizing topological superconductivity, owing to their nontrivial bulk topology and the associated Fermi arc surface states that may arise from broken time-reversal or inversion symmetry \cite{Armitage:18}. Recently, trigonal non-centrosymmetric \pb\ has drawn interest following reports of intrinsic topological surface superconductivity, arising from surface Fermi arcs \cite{kuibarov_evidence_2024,changdar_topological_2025}. 

Bulk superconductivity in trigonal \pb\ has been reported with a critical temperature $T_c = 0.6$--$1.1$~K \cite{Zabala_2024_enhanced_weak}, while surface-sensitive probes such as scanning tunneling microscopy (STM) \cite{schimmel_surface_2024} and angle-resolved photoelectron spectroscopy (ARPES) \cite{kuibarov_evidence_2024, kuibarov_measuring_2025} indicate much higher surface $T_c$. In particular, Ref.~\cite{kuibarov_evidence_2024} reports $T_c = 8$ and $14$~K on the surface Fermi arcs, corresponding to the two different surface terminations, denoted as decorated-honeycomb (DH) and Kagome-like (KL), as explained below. However, photoemission experiments in Ref.~\cite{oleary_topography_2025} do not observe superconductivity down to $3$~K on one of the surface terminations (DH), possibly due to differences in experimental conditions \cite{kuibarov_measuring_2025}. Additionally, the superconducting gap was found to be spatially inhomogeneous \cite{schimmel_surface_2024}, although a more recent work \cite{huang_sizable_2025} finds a homogeneous gap for the DH surface. Other scanning tunneling spectroscopy measurements on the DH and KL surfaces measured at 5 K and 30 mK, respectively, show no sign of superconductivity, despite the presence of Fermi arcs \cite{hoffmann2025fermi}. Moreover, the nature of the superconducting state remains unclear: Ref.~\cite{changdar_topological_2025} proposes $i$-wave gap symmetry on the surface Fermi arcs (KL-termination) based on high-resolution ARPES. Electron--phonon coupling combined with the Fermi arc spin texture has been theoretically suggested as a mechanism for chiral $p$-wave pairing \cite{Maeland_2025_Phonon_mediated_top_supcon}, while including Coulomb interactions may stabilize dominant $i$-wave pairing in certain parameter ranges \cite{maeland2025mechanismnodaltopologicalsuperconductivity, Dsouza2026KL, Buccheri2026ph}.

Given the complex picture emerging from previous results from surface-sensitive probes, a detailed understanding of the bulk and surface electronic structure in \pb\ is of interest. 
Termination-dependent surface electronic states were observed by ARPES and density functional theory (DFT) \cite{Wenxiang_2020_electronic_structure_of_PtBi2}. More recent ARPES works revealed topological Fermi arcs \cite{kuibarov_evidence_2024, kuibarov_measuring_2025, kuibarov_three_2025, changdar_topological_2025,oleary_topography_2025}, in agreement with DFT calculations showing the formation of Weyl nodes close to the Fermi level \cite{vocaturo_electronic_2024, veyrat_berezinskiikosterlitzthouless_2023}, whose origin has been further clarified \cite{Palumbo_2025_PtBi2_peierls, Veyrat2025}. The presence of Weyl nodes in the bulk electronic structure slightly above the Fermi level was also supported in a pump-probe ARPES study \cite{Majchrzak_2025_Machine_learning_PtBi2}. Spin-resolved ARPES measurements provided evidence for Rashba-like spin textures in the bulk electronic structure \cite{feng_rashba-like_2019} and demonstrated spin-momentum-locking for the Fermi arc surface states \cite{mathisen2026}. Earlier ARPES studies have also explored the electronic structure of the centrosymmetric polymorph of PtBi$_2$ \cite{Thirupathaiah_Origin_of_MR_2018, Yao_Bulk_and_Surface_2016}. Despite these previous works, a systematic overview of the termination dependence of electronic states and the distinction between bulk and surface contributions to the complex electronic structure in \pb\ remains incomplete.

Here, we present a detailed study of the bulk and surface band structure for both surface terminations in \pb(0001). Using ARPES at vacuum ultraviolet (VUV) excitation energies in combination with DFT calculations, we disentangle bulk and surface contributions and present a systematic assignment of various features in the valence electronic structure. By varying photon energy and light polarization, we resolve the $k_z$-dependent bulk dispersion as well as termination-dependent surface states. Finally, we analyze the orbital character in the electronic band structure of \pb.

\begin{figure}
    \centering
    \includegraphics[width=1\linewidth]{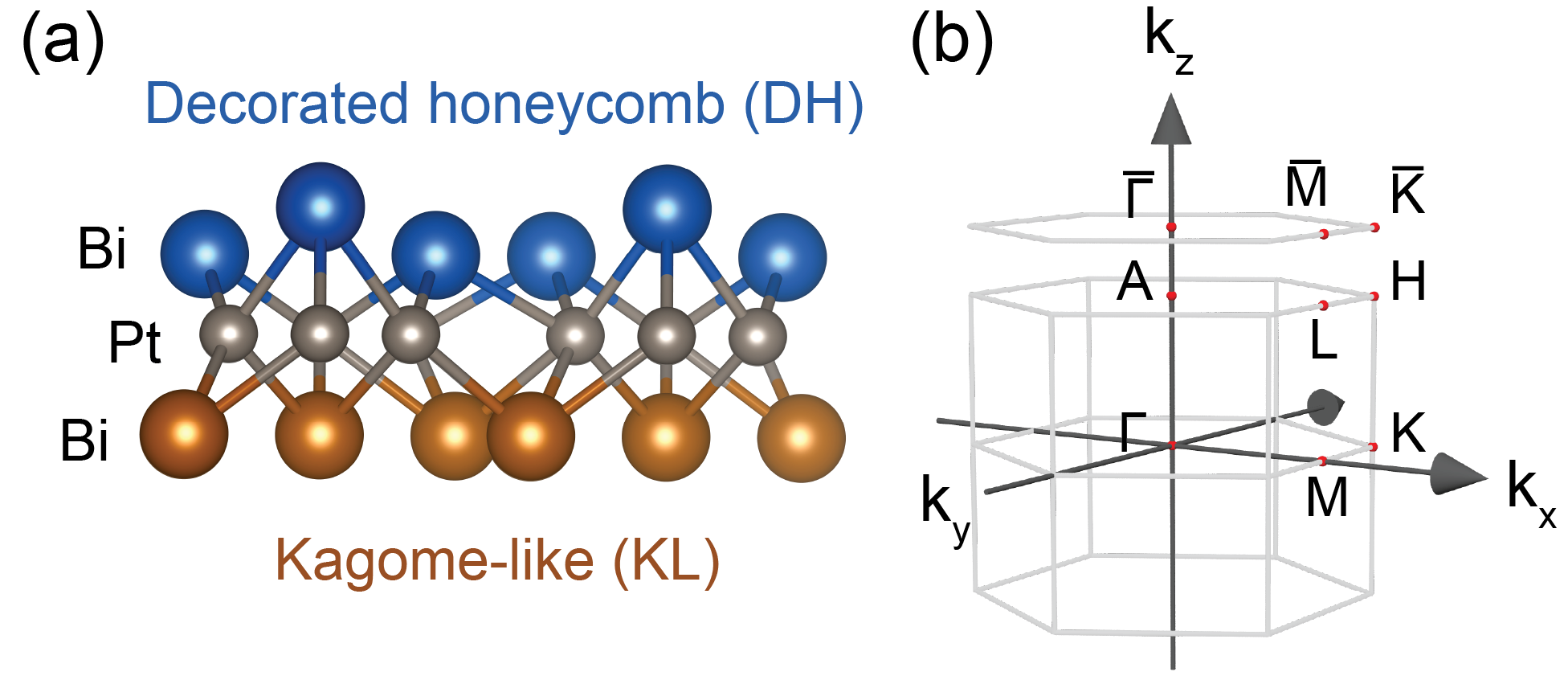}
    \caption{(a) Atomic configuration of trigonal \pb\ where the DH- and KL-terminations are shown in blue and orange \cite{Momma2011}. (b) Hexagonal bulk and surface Brillouin zones of trigonal \pb\ with high-symmetry points indicated.}
    \label{fig1}
\end{figure}

\section{Experimental details}
We carried out ARPES experiments at the Bloch beamline (A-branch endstation) of the MAX IV facility in Lund, Sweden, and at beamline BL-9A of the Research Institute for Synchrotron Radiation Science (HiSOR) in Hiroshima, Japan. Throughout the experiments, the samples were kept at a temperature of around 18 K (Bloch) and 12 K (BL-9A) and in an ultrahigh vacuum below $10^{-10}$ mbar. Data shown in Figs. \ref{fig2} and \ref{fig4} (Bloch) were acquired using a Scienta Omicron DA30-L hemispherical analyzer (slit perpendicular to the plane of light incidence), whereas data in Figs. \ref{fig3} and \ref{fig6} (BL-9A) were acquired using a SPECS ASTRAIOS 190 hemispherical analyzer (slit parallel to the plane of light incidence). The energy resolution was around 20 meV. The range of photon energies used is 18-120 eV (Bloch) and 10-40 eV (BL-9A). ARPES data processing involves centering the images to correct for misalignments, FFT-filtering to suppress the microchannel plate background, and alignment in energy to a common Fermi level. The data were converted to momentum space, with a slight rescaling of the momentum axes to match the Brillouin zone dimensions based on the lattice parameters.

Single crystals of \pb\ with lattice constants $a=b=6.573$~\AA\ and $c=6.162$~\AA\ were grown using a self-flux method as described in Ref. \cite{Shipunov_polymorphic_PtBi2_2020}. Clean surfaces were obtained by \textit{in-situ} cleaving with kapton tape. For a given crystal, repeated cleaves reproducibly yielded the same surface termination. To mitigate surface degradation over time, samples were re-cleaved regularly. 

For the calculations shown in this work, we performed relativistic density-functional calculations (DFT) with  FPLO v22.01-63 ~\cite{Koepernik1999} using the generalized gradient approximation~\cite{Perdew1997} and the experimental crystal structure. Brillouin zone integrations were performed with tetrahedron method  based on a $k$-mesh having $19\times19\times17$ subdivisions. To study the surface electronic structure, we consider semi-infinite slabs of a Wannier Hamiltonian which includes Bi 6$p$ and 6$s$ orbitals together with Pt 6$s$ and Pt 5$d$ orbitals~\cite{koepernik23}.

\section{Results and discussion}

\subsection{Bulk bands}

\begin{figure}
    \centering
    \includegraphics[width=1\linewidth]{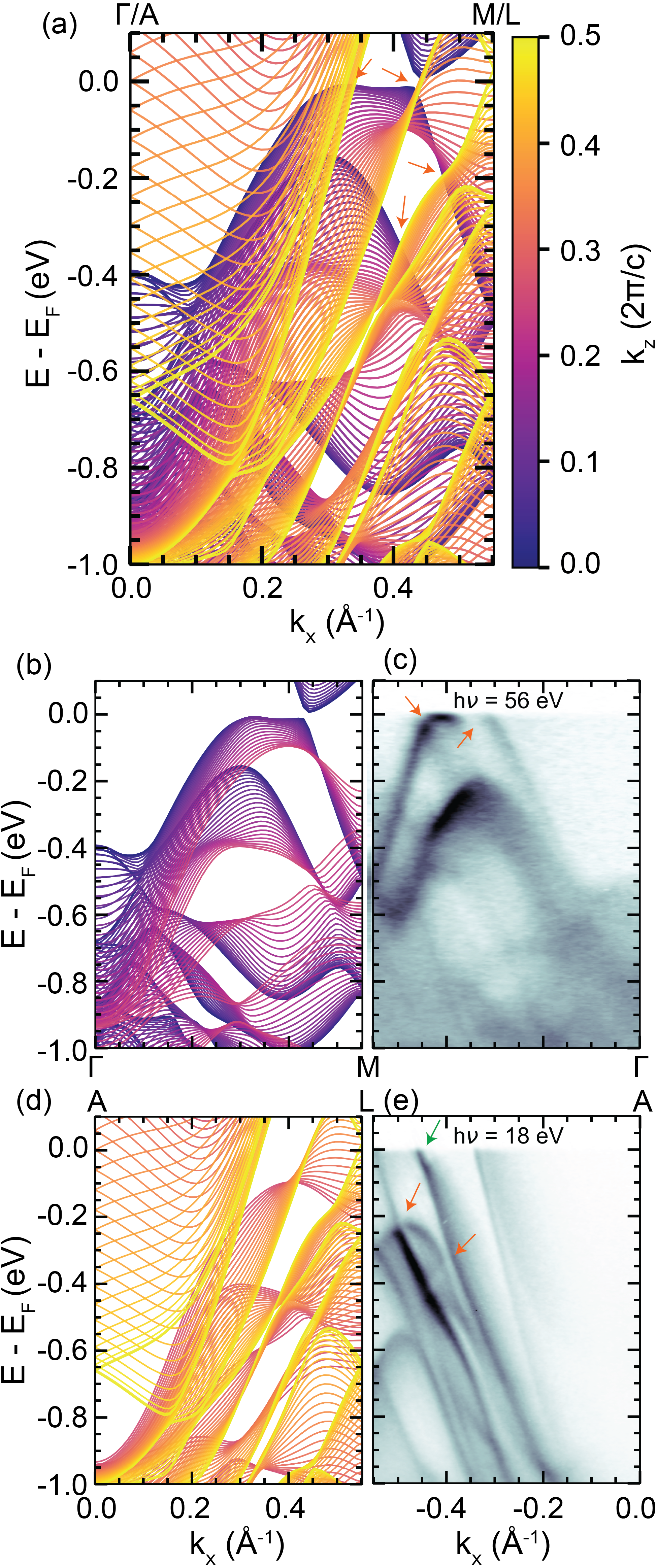}
    \caption{(a) DFT calculation of \pb\ showing the band dispersion for different $k_z$ values along the $\Gamma M$ ($AL$) high-symmetry direction, encoded by color. The calculation is separated into two parts in (b) and (d), highlighting bands near the $\Gamma$-plane and $A$-plane, respectively, while retaining the same color scale as in (a). (c,e) Corresponding experimental ARPES datasets measured on a sample with KL surface termination using $p$-polarized light at photon energies as indicated. Orange and green arrows mark characteristic features, discussed in the text.} 
    \label{fig2}
\end{figure}

In Fig.~\ref{fig1}(a), the crystal structure of non-centrosymmetric \pb\ is shown, which crystallizes in the trigonal space group \textit{P}31\textit{m} (No.~157) \cite{BISWAS_1969_PtBi2,Kaiser_PtBi2_Low_T_reduction_2014,feng_rashba-like_2019,Shipunov_polymorphic_PtBi2_2020, veyrat_berezinskiikosterlitzthouless_2023}. The structure consists of stacked Bi-Pt-Bi trilayers with interlayer van der Waals-interactions. Figure ~\ref{fig1}(b) shows the hexagonal 3D Brillouin zone (BZ) and the corresponding 2D surface Brillouin zone for the PtBi$_2$(0001) surface. The bulk crystal exhibits threefold rotational symmetry about the $c$-axis, and the $\Gamma K$-directions lie in bulk mirror planes. First-principles calculations indicate that the electronic structure near the Fermi level is dominated by Bi $6p$ and Pt $5d$ orbitals \cite{vocaturo_electronic_2024}. The absence of inversion symmetry in the \textit{P}31\textit{m} structure allows for spin spitting in the bulk band structure \cite{vocaturo_electronic_2024}.

We consider the bulk electronic band structure of \pb, using both first-principles calculations and experimental data. In Fig.~\ref{fig2}(a), we present the calculated bulk band structure for different $k_z$ values between the $\Gamma$- and $A$-planes of the 3D Brillouin zone [cf. Fig.~\ref{fig1}(b)], encoded via the color scale. Here, $k_z$ is given in units of $2\pi/c$, such that $k_z=0$ and $k_z=0.5$ correspond to the $\Gamma$- and $A$-planes, respectively. Accordingly, bands shown in purple at $k_z=0$ correspond to the $\Gamma M$ high-symmetry direction in the center of the Brillouin zone, and yellow bands ($k_z = 0.5$) correspond to the $AL$-direction at the top of the Brillouin zone. The orange arrows mark several crossing points between the purple and yellow bands, which will be discussed later.

In agreement with previous calculations \cite{Shipunov_polymorphic_PtBi2_2020, gao_tripple_degenerate_2018, vocaturo_electronic_2024, feng_rashba-like_2019, kuibarov_evidence_2024}, the DFT calculations reveal substantial $k_z$ dispersion, indicating a pronounced three-dimensional electronic character despite the layered crystal structure. For clarity, we further divide the bulk band structure into two $k_z$ regions. Figure~\ref{fig2}(b) shows the states close to the $\Gamma$-plane, covering $k_z = 0$--$0.25$, whereas Fig.~\ref{fig2}(d) displays the region near the $A$-plane with $k_z = 0.25$--$0.5$. In Fig.~\ref{fig2}(b), two bands near the $M$-point exhibit a Rashba-like splitting, studied in detail in Ref. \cite{feng_rashba-like_2019}, with the outer branch located well below the Fermi level. The inner branch starts at approximately $E-E_F \approx -0.5$~eV at the $M$-point ($k_z = 0$) and disperses steeply upward toward $\Gamma$. Moving away from $M$, its group velocity decreases and the band gradually flattens, forming a hole-like dispersion that constitutes the lower cone of the Weyl node (located slightly away from the ${\Gamma M}$ high-symmetry direction). Upon increasing $k_z$, the overall shape of these bands changes only gradually, resulting in a dense bunching of states in energy and momentum space. In the $k_z$-regime closer to the $A$-plane (Fig.~\ref{fig2}(d)), several bands become significantly steeper and display group velocities opposite to that of the inner Rashba-like branch. A similar band bunching of states is observed, although some bands, in particular those around $k_x = 0\ \mathrm{\AA^{-1}}$, have a strong dependence on $k_z$.

In order to compare the experimental data with the $k_z$-resolved calculations, we performed photon-energy-dependent ARPES measurements over the range $h\nu = 18$--120~eV. Figures~\ref{fig2}(c) and (e) show spectra measured on a sample with the KL surface termination. The surface termination should not be of strong influence to the bulk dispersion. Using a photon energy of $h\nu = 56$~eV (Fig.~\ref{fig2}(c)), the observed bands match well with calculations for the $\Gamma$-plane. In particular, both the inner and outer Rashba-like branches around the $M$-point are clearly resolved. As we are measuring in the surface-sensitive VUV regime, due to the finite escape depth, there is uncertainty in the $k_z$ value probed such that each spectrum represents an average over a finite $k_z$-range ($k_z$-broadening). Nevertheless, the Rashba-like bands in Fig.~\ref{fig2}(c) appear sharp. This observation is consistent with the weak $k_z$-dependence of these bunched states as seen in the calculations. Consistently, in the $h\nu$-dependent measurements, the Rashba-like branches remain well defined over a broad photon-energy range rather than exhibiting pronounced energy shifts. 

In Fig.~\ref{fig2}(e), measured with $h\nu = 18$~eV, the observed band dispersion agrees more closely with the calculations near the $A$-plane, consistent with previous studies \cite{kuibarov_measuring_2025, Wenxiang_2020_electronic_structure_of_PtBi2}. Overall, the steep bulk bands in this momentum region are well resolved, including the two bands crossing the Fermi level near $\lvert k_F \rvert \approx 0.34\ \mathrm{\AA^{-1}}$. They remain approximately parallel before evolving towards the crossing at $k_x = 0$, and were previously difficult to distinguish experimentally \cite{Wenxiang_2020_electronic_structure_of_PtBi2}. Closer to the $A$-point ($k_x = 0$) at $E-E_F \approx -1.0$~eV, the spectral intensity is strongly reduced, making the corresponding bands difficult to resolve experimentally. 

\begin{figure*}
    \centering
    \includegraphics[width=1\linewidth]{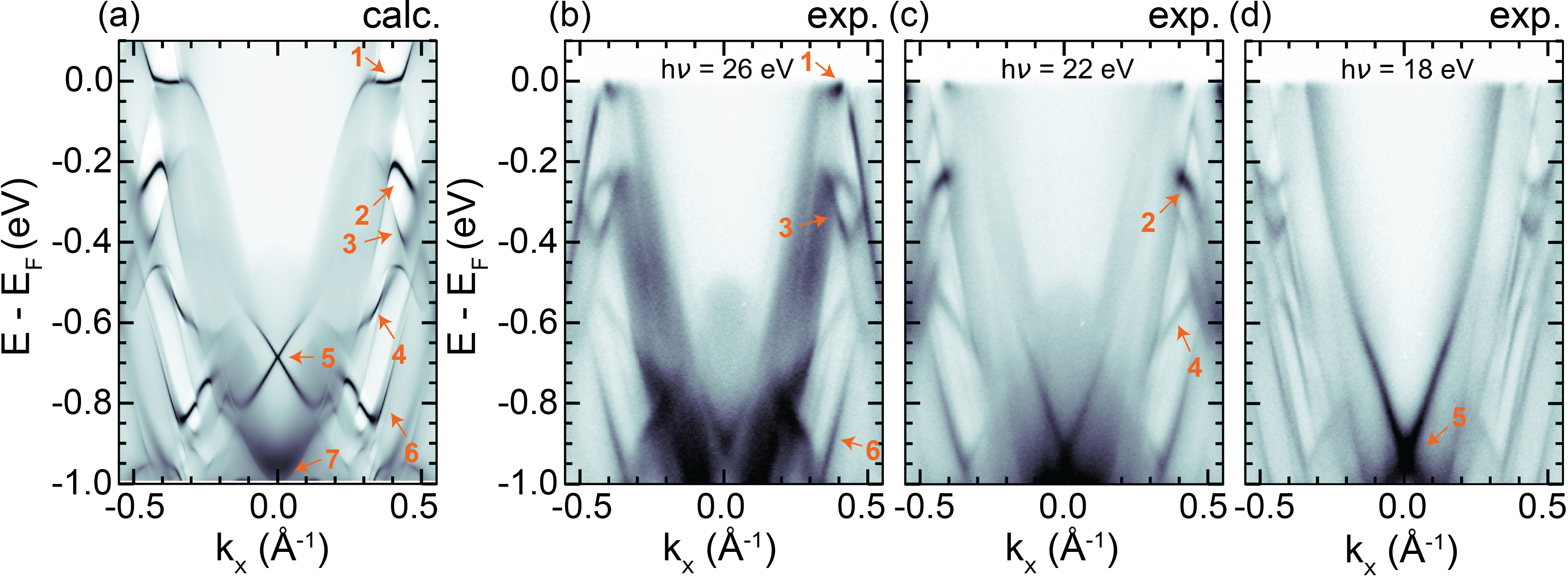}
    \caption{(a) Surface spectral weight calculation of a sample with DH surface termination along the $\bar{\Gamma}\bar{M}$-direction. Prominent surface features are indicated in orange. (b--d) ARPES datasets of the DH surface termination taken at BL-9A, HiSOR, using $s$-polarized light. Experimental features are marked in the panels where they are most clearly resolved.}
    \label{fig3}
\end{figure*}

A closer inspection of the datasets reveals slight local reductions in spectral intensity, highlighted by the orange arrows in Figs.~\ref{fig2}(c) and (e), which we attribute to band hybridization effects. In particular, comparison with the calculated band structure at the two top orange arrows in Fig.~\ref{fig2}(a) suggests that the steep bands in yellow intersect the inner Rashba-like branch at the orange arrows in Fig.~\ref{fig2}(c), which can lead to a suppression of spectral weight \cite{tusche2016multi}. Similarly, in Fig.~\ref{fig2}(e), reduced intensities are observed at the crossing points between the yellow bands and both the inner and outer Rashba-like branches. In Fig.~\ref{fig2}(e), the steep band marked by the green arrow is located where the surface Fermi arc is expected near $E_F$. While it closely resembles the steep yellow bulk band, the surface calculations below reveal its mixed bulk--surface character.

Overall, apart from minor energy shifts, the bulk electronic structure probed by two representative photon energies is well reproduced by the DFT calculations. The complete photon-energy dependent measurements, measured for both surface terminations and for both $s$- and $p$-polarized light, are presented in Fig.~S1 in the Supplemental Material \cite{suppl} as Fermi-energy maps in the $k_x$--$k_z$ plane.

\subsection{Surface states}

\begin{figure*}
    \centering
    \includegraphics[width=1\linewidth]{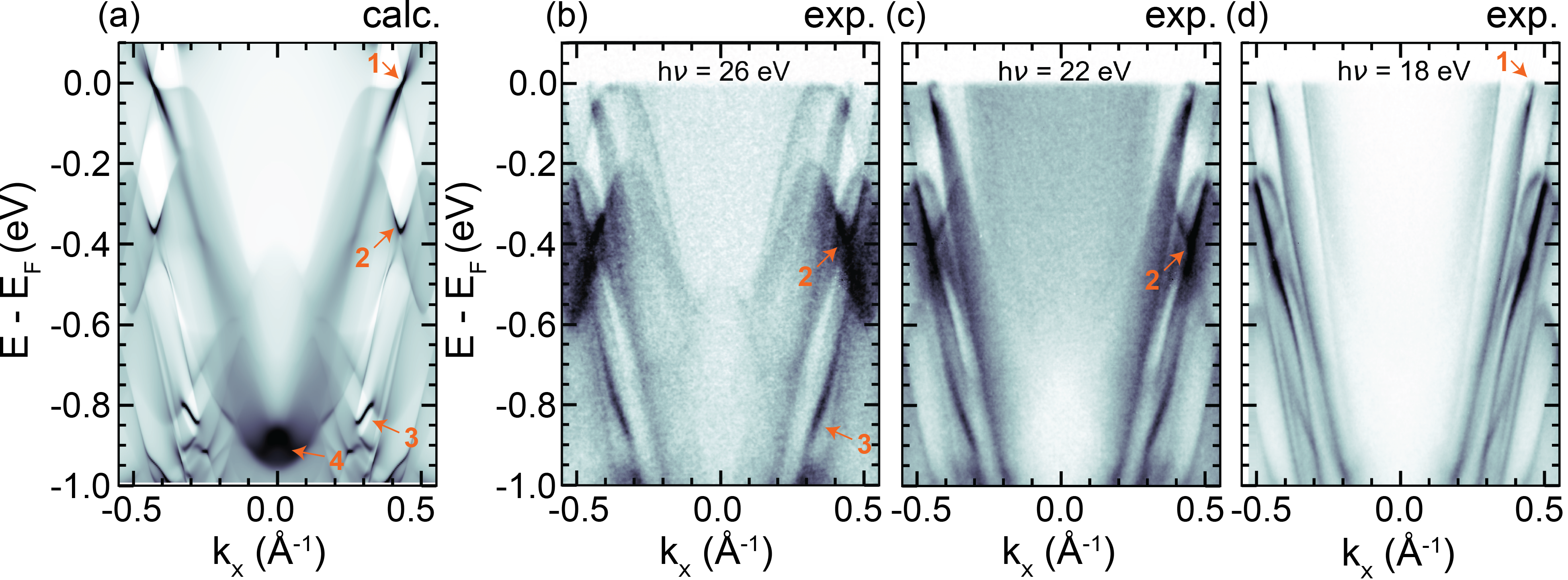}
    \caption{(a) Surface spectral weight calculation of a sample with KL surface termination along the $\bar{\Gamma}\bar{M}$-direction. Prominent surface features are indicated in orange. (b--d) ARPES datasets of the KL surface termination taken at Bloch using $p$-polarized light. Experimental features are marked in the panels where they are most clearly resolved. In contrast to panels (b--d) of Fig.~\ref{fig3}, the analyzer slit is oriented perpendicular to the plane of light incidence.} 
    \label{fig4}
\end{figure*}

We now shift attention to the surface electronic structure.
Two surface terminations can be distinguished for \pb(0001), as shown in Fig.~\ref{fig1}(a). In line with the nomenclature in \cite{vocaturo_electronic_2024}, we refer to these terminations as decorated-honeycomb (DH) and Kagome-like (KL). In the decorated-honeycomb termination, the Bi atoms are corrugated along the stacking direction ($c$-axis), whereas in the Kagome-like termination, the Bi atoms are coplanar. Upon cleaving, only one of these terminations is exposed on a given surface, while the opposing surface of the cleaved crystal exhibits the complementary termination.

In Fig.~\ref{fig3}(a), we show a surface spectral weight calculation along the $\bar{\Gamma}\bar{M}$-direction for a semi-infinite crystal terminating at the decorated-honeycomb (DH) termination. States localized within the surface layers appear sharp and with strong spectral weight, whereas bulk bands (for all $k_z$-values) form a projected continuum, typically with weaker spectral weight. Prominent features are labeled (1)--(7). In Fig.~\ref{fig3}(b--d), we present ARPES measurements of a DH-terminated sample, using $s$-polarized light. The observed spectra depend strongly on the photon energy: using $h\nu = 26$ eV, the data reflect bands from around the $\Gamma$-plane to a larger degree, whereas using $h\nu = 18$ eV we are more sensitive to bands from around the $A$-plane (see Fig.~\ref{fig2}). Surface states, having no $k_z$-dispersion, are in principle present independent of photon energy, although their spectral weight can vary due to matrix-element effects. In Fig.~\ref{fig3}(b), we observe a broad intensity around the $\Gamma$-point at higher binding energies, while the inner Rashba-like band near the $M$-point appears rather sharp. We identify several surface states, including the Fermi arc indicated by (1), which exhibits a relatively flat dispersion near the Fermi level. 

Surface states (3) and (6) are also visible, although state (6) appears at a slightly shifted energy compared to the calculation. Panel (c) resolves surface states (2) and (4) better, while panel (d) shows a pronounced Dirac-like crossing corresponding to feature (5). In the experimental data, this crossing appears at a binding energy of approximately $E - E_F = -0.9$ eV, in line with findings in \cite{Wenxiang_2020_electronic_structure_of_PtBi2, oleary_topography_2025}, whereas it is calculated to lie near $E - E_F = -0.7$ eV. State (7) is likely of mixed bulk-surface character and observable as the broad intensity around the $\Gamma$-point in  panel (b), while its crossing point at $\Gamma$ lies below $E - E_F = -1.0$ eV, shown in Fig.~S2 \cite{suppl}. Apart from the mentioned discrepancies, the surface spectral weight calculations reproduce the experimental data well. 

In Fig.~\ref{fig4}, we present equivalent theoretical and experimental data for the KL surface termination using $p$-polarized light. As in Fig.~\ref{fig3}, photon energies of $h\nu = 26$ eV and $18$ eV preferentially probe bands near the $\Gamma$- and $A$-planes, respectively. In Fig.~\ref{fig4}(a), we observe significant qualitative differences in the surface electronic structure compared to the DH termination. In particular, the surface Fermi arc labeled (1) becomes steeply dispersing and seems to be mixing with the underlying bulk-continuum, as discussed in more detail in Ref. \cite{mathisen2026}. The prominent Dirac-like crossing, seen on the DH termination (feature (5) in Fig.~\ref{fig3}(a)), is absent for the KL termination. In panels (b-–d), a pronounced spectral weight is clearly observed near surface state (2). Surface states (1) and (3) are also present, albeit less readily distinguishable as they are close to bulk states. The broader spectral weight around state (4) is not clearly observed in our data, possibly because the band minimum lies beyond the measured energy range, at higher binding energy than predicted by the calculations. Various surface states can also be distinguished in the equivalent dataset with $s$-polarization, shown in Fig.~S3 \cite{suppl}.

\subsection{Orbital character and polarization dependence}
\begin{figure*}
    \centering
    \includegraphics[width=1\linewidth]{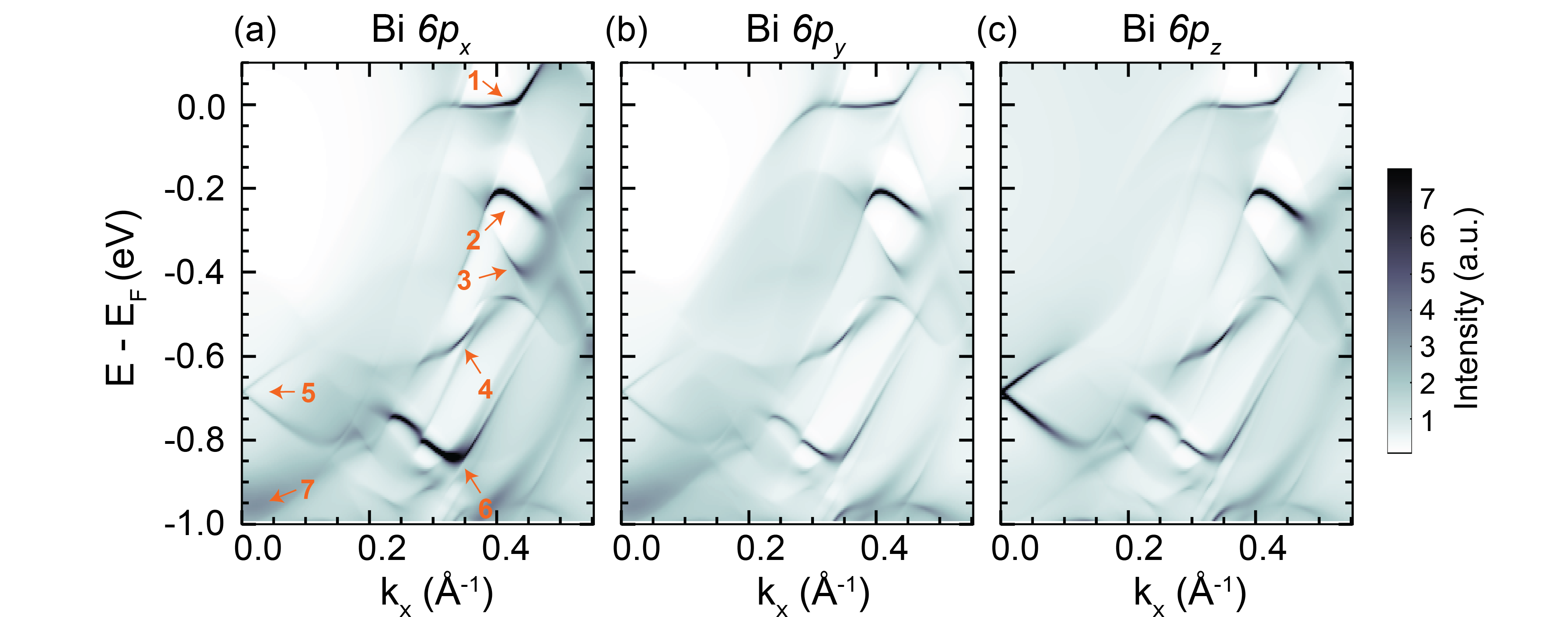}
    \caption{Surface spectral weight calculation of a DH-terminated sample along the $\bar{\Gamma}\bar{M}$-direction, resolved into orbital contributions from (a) Bi $6p_x$, (b) Bi $6p_y$, and (c) Bi $6p_z$. The labels 1-7 correspond to those in Fig.~\ref{fig3}(a). All panels use the same global color scale. The color bar is shown in arbitrary units and serves only as a relative intensity scale. For visualization, the color mapping is clipped to the 1$^{st}$--99.7$^{th}$ percentile of the global intensity distribution. A higher-contrast version is available in the Supplemental Material~\cite{suppl} in Fig.~S6.} 
    \label{fig5}
\end{figure*}
We next examine the orbital character of the surface states in \pb. In Fig.~\ref{fig5}(a--c), we present the surface spectral weight calculation of a semi-infinite crystal with DH termination (the same calculation as shown in Fig.~\ref{fig3}(a)), resolved into the Bi $6p_x$, $6p_y$, and $6p_z$ orbital components. Contributions from other orbitals are comparatively weak in this energy range (in Fig.~S6 \cite{suppl}, we show the calculations resolved for Pt $5d$ and $6s$ orbitals). As shown in Fig.~\ref{fig5}, most surface states exhibit mixed orbital character. For example, the surface states labeled (1), (2), and (6) contain substantial contributions from all three Bi $6p$ orbitals. However, several distinct orbital contributions can be identified. The Dirac-like crossing at the $\Gamma$-point (5) is dominated by the even $p_z$ orbital. Moreover, the broad spectral weight (7) observed around $E - E_F = -1.0$~eV at the $\Gamma$-point originates primarily from $p_x$ and $p_y$ orbitals, with negligible $p_z$ contribution (more clearly visible in the higher-contrast plots provided in Fig.~S6 of the Supplemental Material \cite{suppl}).
\begin{figure*}
    \centering
    \includegraphics[width=1\linewidth]{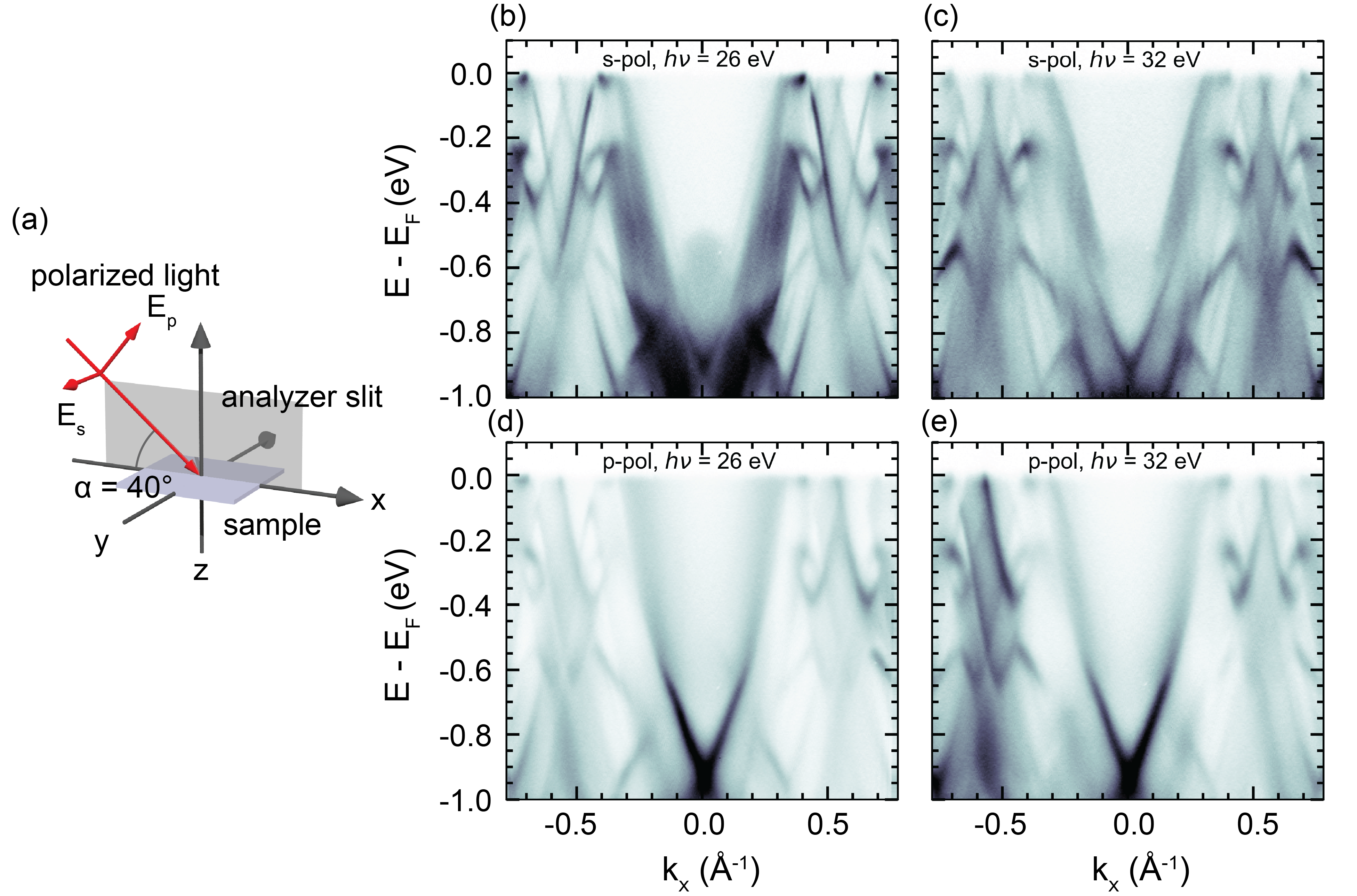}
    \caption{(a) Geometry of the ARPES experiment at BL-9A, HiSOR, showing the definitions of horizontal ($s$) and vertical ($p$) linear light polarization relative to the plane of light incidence. (b--d) ARPES intensity of the DH surface termination measured along the $\bar{\Gamma}\bar{M}$-direction for both $s$- and $p$-polarized light at two photon energies.}
    \label{fig6}
\end{figure*}

The orbital character, as we will explain below, gives rise to characteristic polarization-dependent matrix element effects in ARPES. We therefore first examine the polarization dependence of the spectra measured on the DH termination. Figure~\ref{fig6}(a) illustrates the experimental geometry at beamline BL-9A at HiSOR, where linearly polarized light is incident within the gray plane, parallel to the analyzer slit. The $\bar{\Gamma}\bar{M}$ high-symmetry direction is aligned along the plane of light incidence.

Panels (b)--(e) present ARPES datasets acquired from the DH termination under various experimental parameters as indicated. A comparison of panels (b) and (d), which show spectra recorded at identical photon energies but with different light polarizations, reveals vast differences in spectral features. Under $s$-polarized light, we observe pronounced spectral weight around the $\Gamma$-point at higher binding energies, well-defined sharp Rashba-like bands, and distinct surface states, including states (1), (3), and (6). In contrast, $p$-polarized light significantly suppresses the intensity of these bands while enhancing the visibility of surface state (5). Notably, the steep band crossing the Fermi level at the BZ border ($k_x = 0.55$~\AA$^{-1}$), which corresponds to a bulk state originating from the $A$-plane (see Fig.~\ref{fig2}(d)), exhibits enhanced intensity under $p$-polarization, particularly at the $M$-point with positive $k_x$. This asymmetric intensity distribution along the $k_x$-axis arises from the symmetry breaking induced by $p$-polarized light in the plane of incidence.
\begin{figure}
    \centering
    \includegraphics[width=1\linewidth]{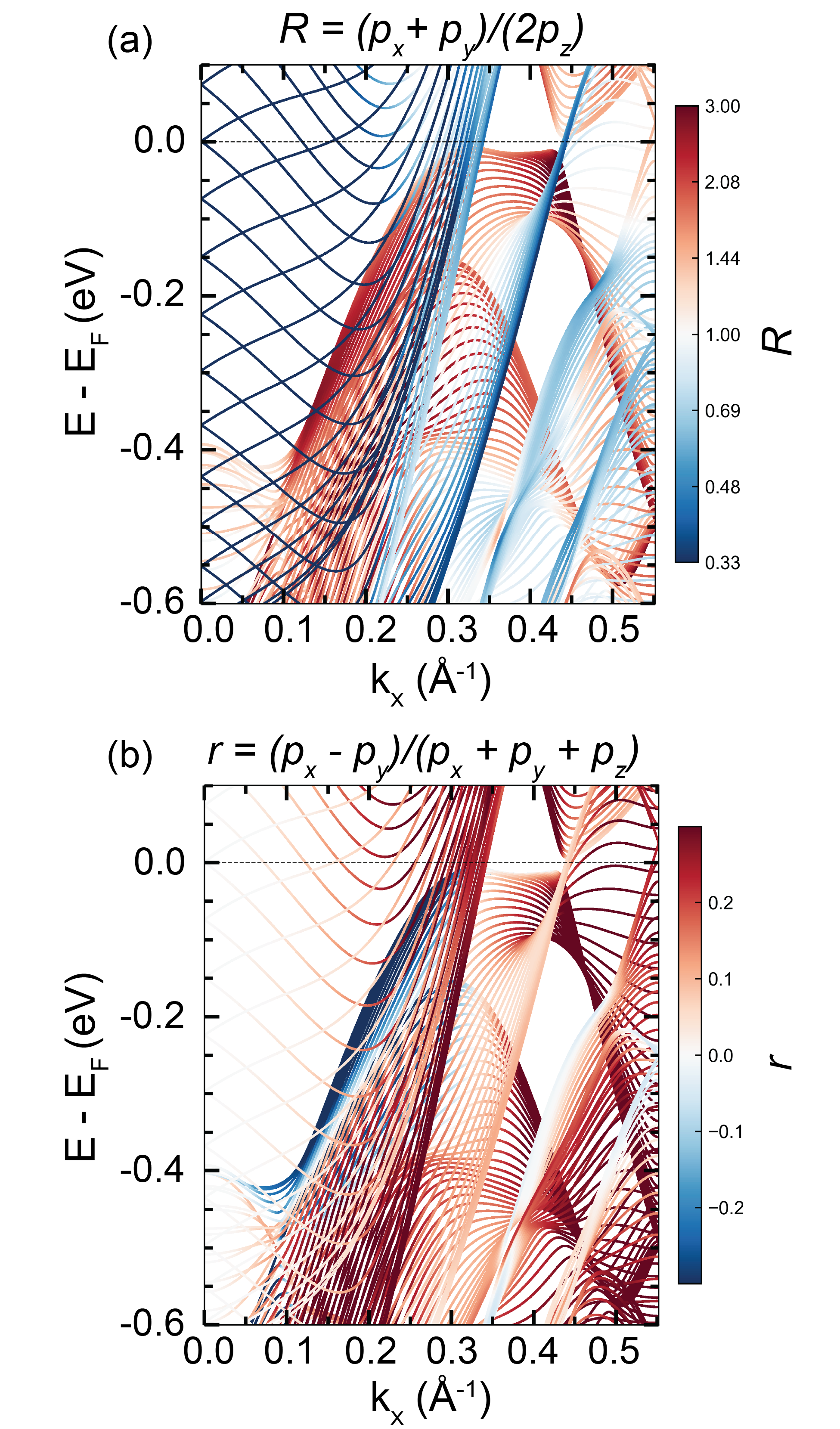}
    \caption{Calculated bulk band structure along $\Gamma M$ for all $k_z$ values between the $\Gamma$- and A-planes. In (a), the bands are colored according to the in-plane to out-of-plane orbital-weight ratio, $R = (w_{p_x}+w_{p_y})/(2w_{p_z})$ (red: $p_x/p_y$ dominant; blue: $p_z$ dominant). In (b), bands are colored by the in-plane orbital decomposition, $r = (w_{p_x}-w_{p_y})/(w_{p_x}+w_{p_y}+w_{p_z})$ (red: $p_x$ dominant; blue: $p_y$ dominant).}
    \label{fig7}
\end{figure}

The datasets acquired at $h\nu = 32$~eV, shown in panels (c) and (e), further demonstrate these polarization-dependent effects. Under $s$-polarized conditions, we observe strong spectral weight on the innermost bands (purple bands in Fig.~\ref{fig2}(b)) around the $\Gamma$-point, along with clearly resolved surface states (2) and (4). However, $p$-polarized light preferentially enhances the purple Rashba-like bands around the $M$-point (particularly for negative $k_x$) and again prominently reveals surface state (5). This polarization-dependent enhancement of surface state (5) under $p$-polarized conditions is consistently observed across the entire range of photon energies explored at BL-9A, HiSOR (10--40~eV), shown in the Supplemental Material \cite{suppl} in Fig.~S5.

The spectral weight observed in ARPES is governed by the photoemission matrix element \cite{Damascelli_2004}, which depends on experimental parameters such as photon energy and light polarization. Within the dipole approximation, the matrix element can be written as
\begin{equation}
M_{f,i}^{\vec{k}} \propto \left| \left\langle \Psi_f^{\vec{k}} \middle| \vec{E} \cdot \vec{p} \middle| \Psi_i^{\vec{k}} \right\rangle \right|^2
\end{equation}
where $\Psi_i^{\vec{k}}$ and $\Psi_f^{\vec{k}}$ are the initial and final state wave functions, $\vec{E}$ is the vector potential of the incident light, and $\vec{p}$ is the electron momentum operator. Although the $\Gamma M$-direction, aligned with the plane of light incidence in our experiment, is not a strict mirror plane of the crystal, dipole selection rules can still provide qualitative insight into the observed intensity variations. Assuming an even final state, the overall matrix element is non-zero (leading to a non-zero photocurrent at the detector) only if the product of the light polarization and the initial state has even symmetry with respect to the plane of incidence. In our geometry, $p$-polarized light $\vec{E}_p = (E_x, 0, E_z)$ is even with respect to the plane of incidence, and therefore predominantly couples to initial states of even orbital character (e.g., $p_x$ and $p_z$). In contrast, $s$-polarized light $\vec{E}_s = (0, E_y, 0)$ is odd with respect to this plane and is thus more sensitive to orbitals of odd character (e.g., $p_y$). These considerations highlight the strong influence of light polarization on the observed spectral weight. In addition, the photon energy affects the properties of the final state, and thereby modulates the photoemission matrix element and the observed spectral weight.

The orbital-resolved calculations for the DH surface termination in Fig.~\ref{fig5} can be related to the polarization-dependent ARPES data shown in Fig.~\ref{fig6}. The datasets measured with $p$-polarized light (Fig.~\ref{fig6}(d) and (e)) display strong intensity at the Dirac-like crossing (5), consistent with the enhanced sensitivity of this polarization to even $p_z$ orbitals. At higher binding energies around the $\Gamma$-point, the strong spectral weight (7) observed in the $s$-polarized data (Fig.~\ref{fig6}(b), and to a lesser extent (c)) can be attributed to the significant contribution from the odd $p_y$ orbital. While the even $p_x$ orbital also contributes in this energy range, its spectral weight is less prominent in the $p$-polarized data, potentially due to the dominant intensity of the Dirac-like crossing. The complete 10–40 eV datasets, acquired in 2 eV steps for both polarizations, are provided in Figs.~S4 and S5 \cite{suppl}. 

We now return to the bulk electronic structure. Figure~\ref{fig7} shows the calculated band structure in a representation analogous to Fig.~\ref{fig2}(a), now with orbital-resolved color encoding. In Fig.~\ref{fig7}(a), the bands are colored according to the ratio $R = (w_{p_x}+w_{p_y})/(2w_{p_z})$, which quantifies the relative weight of in-plane versus out-of-plane orbital character, with large $R$ indicating dominant in-plane ($p_x/p_y$) contributions and small $R$ indicating dominant $p_z$ character. In addition, in order to compare $p_x$ and $p_y$ contributions across the band structure, Fig.~\ref{fig7}(b) highlights the in-plane orbital decomposition via $r = (w_{p_x}-w_{p_y})/(w_{p_x}+w_{p_y}+w_{p_z})$. From Fig.~\ref{fig7}(a), we observe that bands with dominant $p_z$ character (blue) correspond closely to the steep bands originating from large-$k_z$ regions in Fig.~\ref{fig2}(a), consistent with their stronger out-of-plane dispersion. Focusing on the in-plane orbital decomposition in Fig.~\ref{fig7}(b), the inner Rashba-like band (mainly originating from bands around the $\Gamma$-plane, see Fig.~\ref{fig2}(b)) is predominantly $p_x$-like at the $M$-point, but gradually rotates toward increased $p_y$ character as $k_x$ decreases, with $p_x/p_y$ mixing in the nearly flat part of the dispersion. In our experimental data, the Rashba-like states near the $M$-point do not exhibit a clear or systematic polarization dependence across the 10–40 eV photon energy range, although weak variations are observed at selected photon energies. 

\section{Summary}
We have presented a comprehensive study of the bulk and surface electronic structure of trigonal \pb\ using VUV-ARPES experiments in combination with first-principles calculations. The bulk bands exhibit a complex dispersion that we resolved across different $k_z$ planes within the 3D Brillouin zone, revealing features persistent over a broad range of photon energies and highlighting band hybridization effects. 

Comparison of the two surface terminations, decorated-honeycomb (DH) and Kagome-like (KL), demonstrates pronounced termination-dependent differences. In the DH termination, for example, we observe surface-derived states that include a prominent Dirac-like crossing at the $\Gamma$-point, which is strongly influenced by experimental parameters such as photon energy and light polarization. For the KL termination, this Dirac-like crossing is absent. These observations demonstrate that the surface electronic structure is highly sensitive to the local atomic environment, which is determined by the specific surface termination.

Orbital-resolved data for the DH termination provides a connection between ARPES intensity variations and the underlying orbital character, helping interpret the polarization-dependent spectral features. Our results provide a detailed map of the surface and bulk electronic structure, laying the groundwork for understanding surface-related phenomena, including the recently reported topological superconductivity in \pb. Overall, our study provides a framework for disentangling bulk and surface contributions, systematically clarifying termination-dependent electronic properties, and guiding future explorations of topological and superconducting behavior in this material.

\begin{acknowledgements}
\section{Acknowledgments}
This work was supported by the Research Council of Norway through Grant No. 323766 and its Centres of Excellence funding scheme Grant No. 262633 “QuSpin.” We acknowledge the MAX IV Laboratory for beamtime on the Bloch beamline under proposal No. 20250217. Research conducted at MAX IV, a Swedish national user facility, is supported by Vetenskapsrådet (Swedish Research Council, VR) under contract 2018-07152, Vinnova (Swedish Governmental Agency for Innovation Systems) under contract 2018-04969 and Formas under contract 2019-02496. We acknowledge the Research Institute for Synchrotron Radiation Science (HiSOR) at Hiroshima University for beamtime at BL-9A under proposal No. 24AG033. We thank N-BARD, Hiroshima University, for providing liquid helium. JIF thanks Ulrike Nitzsche for technical assistance. F.G. was supported by the Deutsche Forschungsgemeinschaft (DFG, German Research Foundation) through project no. 556350547.

\end{acknowledgements}

\bibliography{references}

\end{document}